\documentclass[conference]{IEEEtran}

\usepackage{graphicx}

\ifCLASSINFOpdf
\else
\fi
\begin{document}
%
\title{Graph Analytics for anomaly detection in homogeneous wireless networks {-} A Simulation Approach}

\author{\IEEEauthorblockN{Nidhi Rastogi}
\IEEEauthorblockA{Department of Computer Science \\
Rensselaer Polytechnic Institute\\
Troy, New York\\
Email: raston@rpi.edu}
\and
\IEEEauthorblockN{James A. Hendler}
\IEEEauthorblockA{Department of Computer Science \\
Rensselaer Polytechnic Institute\\
Troy, New York\\
Email: hendler@cs.rpi.edu}
}


%


\maketitle

\begin{abstract}
In the Internet of Things (IoT) devices are exposed to various kinds of attacks when connected to the Internet. An attack detection mechanism that understands the limitations of these severely resource-constrained devices is necessary. This is important since current approaches are either customized for wireless networks or for the conventional Internet with heavy data transmission. Also, the detection mechanism need not always be as sophisticated. Simply signaling that an attack is taking place may be enough in some situations, for example in NIDS using anomaly detection. In graph networks, central nodes are the nodes that bear the most influence in the network. The purpose of this research is to explore experimentally the relationship between the behavior of central nodes and anomaly detection when an attack spreads through a network. As a result, we propose a novel anomaly detection approach using this unique methodology which has been unexplored so far in communication networks. Also, in the experiment, we identify presence of an attack originating and propagating throughout a network of IoT using our methodology.
\newline
\end{abstract}
\section{Introduction}
Denial of Service (DoS) attacks have been identified as the top threat and IoT are increasingly becoming a powerful tool for attackers engaged in DoS \cite{IEEEhowto:kaspersky}. DoS is a kind of intrusion that renders certain services inaccessible by the legitimate user. It can be detected using a network intrusion detection system (NIDS) deployed on the network at strategic junctions. These NIDS can be broadly categorized into two. Firstly - misuse detection, where the malicious behavior of the network has been identified from prior attacks and have been cataloged for future detection. Secondly- anomaly detection, where behavior outside of modeled baseline behavior is termed as anomalous. The fundamental difference between these two categories lies in what information are they seeking from the collected system data. While misuse-detection searches for description that is close to a known malicious behavior, anomaly detection works on the notion of separating normal behavior from the rest. In anomaly-detection, normal network activity is identified and any deviation is flagged. As such, these systems also do not burden the network with heavy post-data collection analytic work. This is true at least as long as the network is not very dynamic forcing the baseline model to be recorded and compared frequently. This is the reason why we lean towards anomaly detection.
\newline
At present, anomaly-detection systems are heavily reliant on statistical or machine learning techniques, although its effectiveness has sometimes been questioned. \cite{IEEEhowto:sommer}. The author argues that machine learning is most effective in anomaly-detection when the misuse activity is similar to something previously seen, alibi the exact definition of the activity may not be needed. This means that the system needs to be fed a-priori knowledge on the basis of which it can draw conclusions and identify a data-set as anomalous. This may not always be possible. As newer attacks are increasingly coming to the forefront, any prior knowledge of them may not exist.
\newline
This motivates us into exploring graph-based anomaly detection. The main advantages that this method over any other are: (a) networks represented as graphs can be explored for their correlations over a long-period of time. (b) a graph entity is not analyzed in isolation, but in correlation to other entities, (c) multi-dimensional characteristics can be analyzed at a singular level as well as in clusters, both big and small.
\newline
Within graph based network data analysis, we explore information centrality (IC) as a method to identify central nodes from a given network. These nodes are important because of the way they are identified as central. IC uses the location and connectivity of nodes in a given graph to identify those that are quickly accessible. This makes them quicker to respond or identify any change in network behavior. We use this property of information centrality to identify important nodes, followed by using these nodes to detect anomalous behavior in the network.
\newline
Anomaly detection mainly depends on how the detector is built and what needs to be detected. Same detection principle cannot apply for all kinds of attacks. A spear phishing attack targets one clearly identified user (called spear phishing) has a different attack signature as opposed to a denial of service attack or malware targeted towards a larger organization. An attack signature is a unique way of identifying a pattern of information that describes an attempt to exploit a system. In this research, we identify volume anomalies, a class of anomaly that affects a large part of the network. Examples of attacks where this kind of anomaly detection can be useful include: DoS (Denial of Service) attacks, unusual end-user file transfer, etc. 
\newline
The key approach is described in three parts: First, it is demonstrated that Information Centrality (IC) can condense a large graph representing an IoT network  – called graph-sparsification – by identifying nodes that may prove to be central in exhibiting the overall network behavior. This is a critical step in reducing computational complexity associated with large networks since focus is on fewer nodes. Second, the approach is substantiated by simulating a mid-sized ZigBee network (200 nodes) using NS-2 over a mesh topology. The packet transfer uses constant bit rate (CBR) over an AODV routing protocol. Third, a large quantity of packets is introduced that can lead to detecting volume anomaly in the network. By this we mean, the anomaly impacts a large part of the network and is detected by several entities. Through various simulation experiments, we evaluate the ability of central nodes to identify anomaly. We change the data packet size when it is introduced as an anomalous occurrence in the network. Although, this approach is optimized for deployment on constrained nodes with limited energy and memory capacity but can be extended to other networks.

\section{Literature Study of Anomaly detection}
Intrusion detection in networks identifies the presence of an internal or external miscreant that should be quarantined or removed at the earliest. For this, NIDS tools have been created using distinguishable technologies, anomaly detection being one of them. As mentioned earlier, using this technique, systems identify deviations from expected behavior based on the description of normal system activity \cite{IEEEhowto:denning}. 
\newline
Chandola et. al \cite{IEEEhowto:chandola} extensively cover various anomaly detectors as well as the application domain and knowledge disciplines they are developed for. Machine learning offers a wide range of tools for today's systems, such as neural networks, support vector machines, etc. However, \cite{IEEEhowto:sommer} expresses concern that machine learning is rarely employed in practice because of the large variation of benign traffic in operational real world setting. While this may not hold true anymore, as many current security systems have successfully deployed machine learning techniques, a key take away from this paper was that a sound understanding of the system that needs protection is needed.
\newline
We are also of the opinion that the most critical requirement for creating a relevant tool for any environment is to gain deep insights into the system functioning, capabilities, and limitations. Treating one like a black-box and expecting the detection tool to work instantly, will lead to erroneous results and very high false-positives. A quick fix to this is to increase the sensitivity of the tool. But one cannot guarantee this will remain an effective strategy over time.
\newline
\section{Graph Based Analytics}

Large networks send enormous amount of data from one device to another in order to exchange information. For this, they identify routes that have the shortest overhead to the network, and yet can reliably communicate with various nodes. Aside from scaling issues due to incredible number of devices and their connectivity, understanding the behavior of the network at a given time is very challenging and poses a series of problems. Results are generated after analyzing data that not only comes from the connected devices, but also links that connect them and other network managing entities.
\newline
The success of a network monitor is thus crucially dependent on how timely can it report metrics that are useful in understanding the behavior of the network, including any noteworthy changes. These changes can direct the network analyst towards issues that can affect the network and its users. Contemporary network monitors use intrusion detection systems, firewalls, network and device event logs, or a combination of these for this purpose. Besides straightforward techniques that identify deviation from the threshold of network measurement attributes, there are other statistical techniques that have been successfully deployed.
\newline
Graphs can be used to model a wide variety of structures and relationships, like networks, and give us terminologies to explain them. Since graphs have the ability to visualize a system at a higher level, they give us insights by asking questions that can help develop the solution. For example, during the course of coming up with an approach, a researcher is asked pertinent questions such as - what will be the size of the graph, will it be dense or sparse, which algorithms will be most suitable for a graph with a given structure and attributes, will the graph be dynamic or static, etc. Although these questions appear general purpose in the beginning, they indeed provide deep insights into the problem and possible approaches.
\newline
In this section, we examine the structural properties of networks and briefly review relevant properties using graph theory. Consider a graph consisting of a vertex for each device, and a directed edge from a vertex if the corresponding device contains a network link in that direction to the other one. This graph is called the network graph and represents the link structure of devices on the network. Since a link corresponds to communication channel over which data transfer from one node to another in a given direction, it embodies the idea that it contains relevant information. And that the node plays a relevant role in transferring this data to its final destination. And, the node is able to do that precisely because of its location and connection to other nodes. The weight of all the edges is constant across the graph, and hence does not play a differentiating role for a particular edge. Thus, it is reasonable to assume that the structural relevance of a node with respect to other nodes in the graph is called the information centrality of the node. Like all centrality measure, IC is also a relative metric and does not find an absolute value.
\newline
Using links, the graph shows connections between various entities, like people or web pages, which in turn are represented by nodes.

Similar estimates of important vertices in a graph have been proposed earlier, but they all apply to other application domains. No comprehensive, and accepted usage scenario is available in literature. This is precisely what makes this research unique and an attempt to break new grounds in the field of communication network analysis.

\subsection{Centrality Measure and Usage Paradigm}
The concept of centrality indices, or just centrality, originally came in the late 1940s from studying human communication in small groups \cite{IEEEhowto:bavelas}. Soon centrality gained a strong foothold as an estimate of an individual's importance in social networks \cite{IEEEhowto:freeman}. It established relationships between particular structural features of a network and an actor's influence. In other words, it identified "the most important actor" or "the most useful entity" in a social graph (where a vertex is an actor), alibi using ad-hoc formalization. With time, many centrality measures were introduced, which were based on the link between pair of vertices, namely- degree, betweenness, closeness, eigenvector, information \cite{IEEEhowto:borgatti}, etc. These measures and linkages between actors were used in several network-analytic studies to evaluate fairly large networks successfully \cite{IEEEhowto:brandes}. Nevertheless, the focus of research remained on social networks, community organizations, and planning for the most part. However, recent conceptual work identifies the suitability of centrality measures in communication networks \cite{IEEEhowto:koschutzki}. It exhibits a close relationship between flow of current in an electrical network and random walks around a graph. These results served as one of the key stimuli behind using centrality measure in communication networks.
\newline
Centrality measures also suffer from a major drawback. When not used for the appropriate network, or flow, they can often lead to incorrect understanding of results. If the centrality does not relate to the purported index, it becomes difficult to understand the measurement. Hence, a clear understanding of the concept of centrality type and its potential for an application is very important. For example, degree centrality counts the number of ties incident upon a node and is not the correct choice for identifying central nodes based on information flow. 
\newline
Each measure identifies central nodes depending on the type of network and connections, as well as the specific feature that defines it. Here, we only describe Information Centrality in detail. We provide an intuitive rationale and an empirical demonstration of the utility of Information Centrality.
\newline
\textit{Information Centrality (IC)} measure is based on the “information” contained in all possible paths between pairs of nodes in a graph \cite{IEEEhowto:stephenson}.  IC has not been explored extensively so far for any kind of network. Its understanding has therefore, been quite vague in terms of applicability, and usage methodology. In this research, we've extended the usage of IC for the first time and demonstrated it empirically through simulation based experiments. The reason we consider this metric is because of two primary reasons. Firstly, no other metric can be used as they cover different aspects of a graph that have no relation to data flow, node location, or time taken by packets to travel between nodes.
\newline
Earlier research uses IC on networks represented by non-directional graphs in which information is important and central nodes are found. However, in all the previous work, there are several limitations. They consider all paths between a pair of nodes and use that to identify central nodes. They use non-directional networks, which is not how real networks exist. None of the previous work demonstrate the usage of IC in communication networks \cite{IEEEhowto:everett}.
\newline
Betweenness Centrality refers to identifying central nodes based on the frequency with which a node lies between pairs of source and destination nodes in the network. In communication networks, the betweenness centrality is a measure of the degree that each node lies between other nodes, transfers packets and thereby gains a sense of importance in contribution to a solution. The higher the frequency of the packets transferred through a node to other nodes, the higher the betweenness centrality, which in turn makes the node participation very important the network.
However, with information centrality, the centrality of the node is defined by using the harmonic average. A major difference between how we use IC and its original definitions is that we do not use all paths to calculate the information measure for each node. We instead use the IC formula along routes that are defined as geodesics and paths only \cite{IEEEhowto:jackson}. The information measure,between two nodes is defined as the reciprocal of the topological distance between the corresponding nodes\cite{IEEEhowto:stephenson}\cite{IEEEhowto:amrit}, 
\[I{ij} = 1/d{ij} \]
Since, reciprocal of the inter node distance is the reciprocal of data transfer, the information between two nodes is the inverse of this inter-node distance. When several routes can transfer data between the same pair of source-destination nodes, the information is calculated considering all these paths. In the real world of communication networks, these routes are mostly shortest paths(geodesics) or the next best option if the shortest path is unavailable(paths). Hence, we use these two distance measuring options to calculate the IC of a node that fall between various pairs of source-destination nodes.

Weight between nodes or bandwidth is kept same between nodes as that too would affect the choice of central nodes \cite{IEEEhowto:stephenson}.

\section{Experiment}
\subsection{Motivation}
The goals behind simulating a communication network and using information centrality to identify central nodes are: Firstly, to see if the simulations point to the same central nodes as when calculating the information centrality. And secondly, to see if the central nodes can indeed identify anomalous behavior earlier than the combined data from all the nodes from the network.
\subsection{Network Simulator}
We use a popular network simulation platform called network simulator 2(NS2)\cite{IEEEhowto:mccane}. It is an open source event-driven simulator widely used in the research community to simulate computer communication networks\cite{IEEEhowto:issariyakul}. It contains several modules for numerous network components, which can be studied using the extensive literature available online for free.
\subsection{Simulation Environment}
The setup chosen for the simulations is based on the scenario where several sensor nodes are connected to various devices in and out of reach environment. These sensor nodes store data from the surroundings and forward them to a central, more powerful node, which in turn sends it to a server in the cloud.
\newline
We choose ZigBee technology to simulate this setup as it is a low cost, low power consumption, and short distance wireless communication technology. It was developed for wireless personal area network (WPAN) in early 2000. The main advantages of simulating ZigBee is that it is widely used in the building automation control, monitoring and control of IoT. It gives us a relatively simpler environment to test our hypothesis on, as well as is much easier to gain expertise in. As mentioned earlier, domain expertise is the top criteria for analyzing any system effectively. We use mesh network topology for ZigBee devices as it allows full peer-to-peer communication and routing is decentralized. Ad-hoc on demand distance vector (AODV) routing is used. We keep the nodes static as this prevents routes from changing frequently. This further allows us to track packets, follow their routing thus making it easier to demonstrate and prove our approach. This, however, in no way is a limitation, and is one of the topics for future work of our research. 
\newline
\newline
Table \ref{table:1} summarizes the general simulation parameters.

\begin{table}[ht!]
\centering
\begin{tabular}{ l | r }
  \hline			
  Parameter & Value \\
  \hline
  Number of nodes & 200 \\
  Simulation Area & 100x100 \\
  Total simulation time & 900 s \\
  Number of simulations & 100 \\
  Traffic Type & CBR \\
  Queue Model & Queue/DropTail/PriQueue \\
  Max. packets in Queue & 1000 \\
  Packet rate & 2/ms \\
  Packet Size & 0.5KB \\
  Total connections & 35 \\
  \hline  
\end{tabular}
\newline
\caption{Simulation Parameters}
\label{table:1}
\end{table}

\subsection{Method}
Once the experimental setup is ready, the NS2 simulator is run using the code written in Tcl/OTcl language. The location for each node is randomly calculated using a python script. The simulation can be visualized using Nam visualization tool, which is part of NS2. A post simulation trace file is generated after the tcl script is run and is used as an input to Nam. We wrote another python script to analyze and comprehend the trace file. It is using this data set that we calculate the arrival time of packets on various nodes for all packet transfers. This is a crucial part of our research. By recording the time it takes for a packet to reach a node, we can identify those nodes that are the ones that are reached in the shortest time. This is what makes them central to the network. 
\newline
We repeated this simulation hundred times for various combination of source and destination nodes, thereby simulating the flow of data packets in the network and ensuring that we get a similar set of nodes as central nodes. The topology remains the same for all the simulations. The multiple number of simulations ensured that we covered all possible data transfer routes and got similar results. The route taken by the packets followed the route defined by AODV routing protocol which we described in earlier section.
\newline
AODV is a reactive or on-demand protocol and discovers routes as and when there is a need to transfer packets between nodes. Its ultimate goal is to avoid duplication or re-sending the same packet, which is accomplished by keeping a sequence number for preventing loops and keeping the fresh most route. Each node maintains its monotonically increasing sequencing number which increases every time the node notices change in neighborhood topology. So when a route request is sent from a node towards the nodes that possibly fall on the path of destination node, the routing protocol makes sure that no loops are encountered. In graph theory language, this means that trail and walks cannot be used to find distance between pairs of nodes. This leaves us with two options - geodesics and path. Geodesic can be called the distance with the shortest route, and path can be any route between nodes where no node or link is repeated.
Once we had average arrival times of all nodes, we chose the top 15\%-20\% nodes when ranked in ascending order of average arrival time. It is to be noted here that we compared the results of central nodes calculated using our simulations and the information centrality measure.
We further use this result to understand the role played by these central nodes in the networked graph. \newline
The goal of our research is to find novel ways of detecting intrusions in communication networks by detecting anomalous behavior in these networks. Our next step post identification of central nodes is to use them for anomaly detection. We hypothesize that since the central nodes are crucial to the flow of information in the network, they should also be central to flow of anomaly in the network. Here it is important to remind our readers that anomalies can be of different categories. Since IC is calculated based on the flow of information in an entire network, same holds true for the anomaly. Our initial intuition is that central nodes can help identify attacks and its related anomaly, which spreads across a network. Example of such an attack is denial of service (dos) and a possible related anomaly could be excessive ping commands on a certain network port.
We use the same central nodes identified earlier for our simulated network and monitor the behavior of these nodes when an anomaly is introduced into the network.
\newline
\begin{table}[ht!]
\centering
\begin{tabular}{ l | r }
  \hline			
  Parameter & Value \\
  \hline
  Normal packet size & 0.5Kb \\
  Anomalous packet size & 10Mb, 50Mb \\
  Anomaly origin & randomly selected 5\% nodes  \\
  Central nodes chosen & top 15\%, 20\% \\
  Anomaly injected & 80th sec \\
  \hline  
\end{tabular}
\newline
\caption{Anomaly Parameters}
\label{table:2}
\end{table}
\subsection{Results and Analysis}
The goal of this experiment is to demonstrate the impact central nodes can have when employed for anomaly detection. To this end, we monitor the behavior of all nodes, including central nodes and capture their effectiveness in detecting anomaly. We keep a track of the time when the anomalous behavior is identified w.r.t the anomaly injection time. We then compare the percentage of central versus non-central nodes that have identified the anomaly since injection. See Figure 2 below for a comparison.

\begin{figure}[h!]
 \centering
\includegraphics[width=0.18\textwidth]{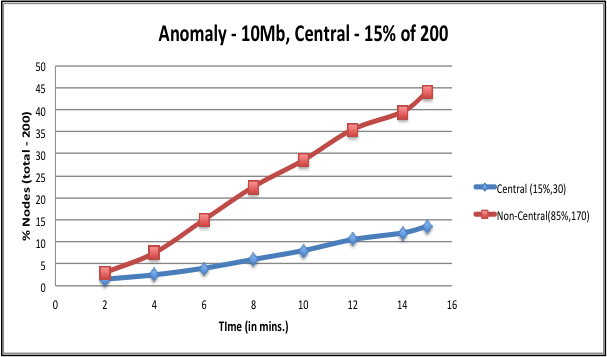}
\includegraphics[width=0.18\textwidth]{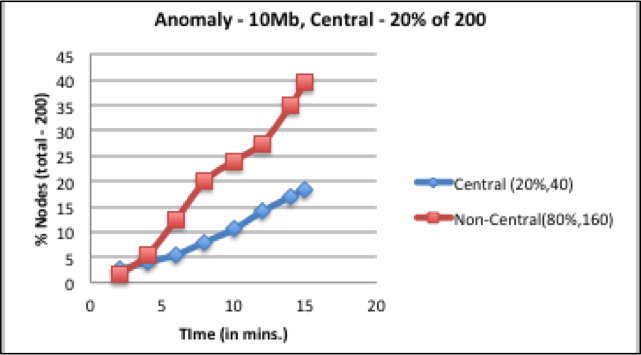}
\includegraphics[width=0.18\textwidth]{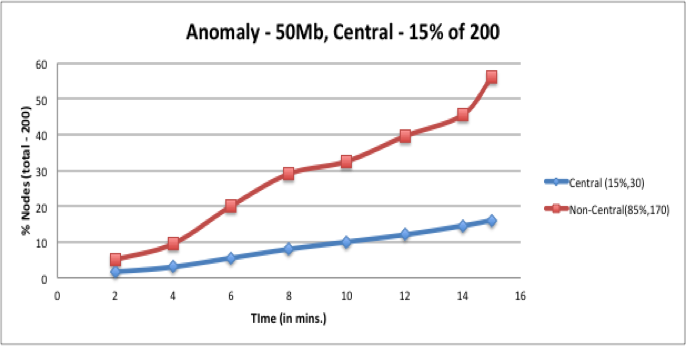}
\includegraphics[width=0.18\textwidth]{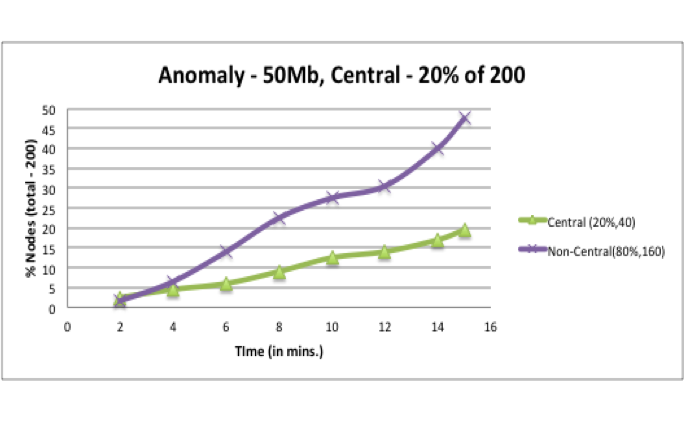}
\caption{Comparison of detection times of central nodes vs. other nodes.}
  \end{figure}

In this research, we assume that the anomaly detector is fairly efficient and do not delve deep into the false rate of the detection mechanism. This is being further analyzed in our future research.

The results show us that for an Anomaly of size 10MB, and monitoring top 15\%-20\% of central nodes, as the anomaly propagates into the network the central nodes are able to detect it’s presence faster than the rest of the nodes. Monitoring the central nodes is a useful exercise in comparison to monitoring the entire network as approximately 90\% of central nodes were aware of the anomaly while only approximately 50\% of non-central nodes for the duration of the simulation.

For an Anomaly of size 50MB, and monitoring top 15\%-20\% of central nodes:
As the anomaly propagates into the network the central nodes are able to detect it’s presence faster than the rest of the nodes
Monitoring the central nodes is a useful exercise in comparison to monitoring the entire network as approximately 99\% of central nodes were aware of the anomaly while only approximately 60\%-65\% of non-central nodes for the duration of the simulation.

Another advantage of using central nodes is the lack of reliance on special hardware or devices dedicated to intrusion detection. The central nodes a wholly part of the network and can accomplish the detection part without dedicating a lot of memory and procession power to the task.

\section{Conclusion}
The main contributions of this research include proving that Information Centrality can be used in communication networks for identifying important nodes. These nodes are significant because of the control they have on the flow of information. IC identifies these nodes based on their location and connectivity to other nodes using a harmonic average of the distance. The simulation compares these values by calculating the average arrival times of data from other nodes along the routes identified in graph theory as geodesics and path. Because of this reason, anomalies, like regular data, also travel past these central nodes and can be identified earlier. This is in comparison to the analysis provided by running an IDS through all the nodes in the network and how quick can it identify the presence of a volume anomaly in the network. We do make it clear that the focus of our research is on volume anomalies that spread across the network and are not targeted attacks or intrusion efforts.




\begin{thebibliography}{1}

\bibitem{IEEEhowto:kaspersky}
Kaspersky DDOS intelligence report for Q3 2016. https://securelist.com/analysis/quarterly-malware-reports/76464/kaspersky-ddos-intelligence-report-for-q3-2016/

\bibitem{IEEEhowto:sommer}
R. Sommer and V. Paxson, “Outside the Closed World: On Using Machine Learning for Network Intrusion Detection,” 2010 IEEE Symposium on Security and Privacy, 2010.

\bibitem{IEEEhowto:denning}
D.E. Denning, "An intrusion-detection model". IEEE Transactions on software engineering, pp. 222-232, 1987.

\bibitem{IEEEhowto:chandola}
V. Chandola, A. Banerjee, and V. Kumar, “Anomaly detection: A survey,” ACM Comput. Surv., vol. 41, no. 3, pp. 1–58, 2009.
\bibitem{IEEEhowto:borgatti}
S.  Borgatti and M.  Everett, "A Graph-theoretic perspective on centrality", Social Networks, vol. 28, no. 4, pp. 466-484, 2006.
\bibitem{IEEEhowto:bavelas}
A.  Bavelas, "A Mathematical Model for Group Structures", Human Organization, vol. 7, no. 3, pp. 16-30, 1948.
\bibitem{IEEEhowto:freeman}
L.  Freeman, "Centrality in social networks conceptual clarification", Social Networks, vol. 1, no. 3, pp. 215-239, 1978.
\bibitem{IEEEhowto:stephenson}
K. Stephenson and M. Zelen, “Rethinking centrality: Methods and examples,” Social Networks, vol. 11, no. 1, pp. 1–37, 1989.
\bibitem{IEEEhowto:brandes}
U.  Brandes, "A faster algorithm for betweenness centrality*", The Journal of Mathematical Sociology, vol. 25, no. 2, pp. 163-177, 2001.
\bibitem{IEEEhowto:koschutzki}
D. Koschützki, K. A. Lehmann, L. Peeters, S. Richter, D. Tenfelde-Podehl, and O. Zlotowski, “Centrality Indices,” Network Analysis Lecture Notes in Computer Science, pp. 16–61, 2005.
\bibitem{IEEEhowto:issariyakul}
T. Issariyakul and E. Hossain, Introduction to network simulator NS2. New York: Springer, 2012.
\bibitem{IEEEhowto:mccane}
S. McCanne, S. Floyd, K. Fall, and K. Varadhan, "Network simulator ns-2," pp. 1059-1068, 1997.
\bibitem{IEEEhowto:amrit}
A. Chintan, J.Hillegersberg, and K. Kumar, "Identifying coordination problems in software development: finding mismatches between software and project team structures," arXiv preprint arXiv:1201.4142, 2012.
\bibitem{IEEEhowto:everett}
E. M. Rogers, and D. L. Kincaid, "Communication networks: toward a new paradigm for research," 1981.
\bibitem{IEEEhowto:jackson}
M. O. Jackson,Social and economic networks Vol. 3. Princeton: Princeton university press, 2008.
\end{thebibliography}
%

\end{document}